\documentclass[oneocolumn,preprintnumbers,amsmath,amssymb]{revtex4}

\usepackage{amsmath,latexsym,amssymb,amsfonts}
\usepackage{float} 
\usepackage{graphicx}
\usepackage{epstopdf}
\usepackage{graphics}
\usepackage{caption}
\usepackage{subfigure}
\usepackage{bm}
\usepackage{color}
\usepackage{footnote}
\usepackage[dvips]{psfrag}

\usepackage{soul}  
\usepackage{xcolor}
\setstcolor{red}

\begin{document}

\title{Quantum Suppression of Mass Inflation in Reissner–Nordstr\"{o}m Interiors via Wheeler–DeWitt Equation}

\author{
Chen-Hsu Chien$^{a}$\footnote{{\tt chien@math.muni.cz}},
Woosung Song$^{b}$\footnote{{\tt thddntjd06@naver.com}},
Gansukh Tumurtushaa$^{c}$\footnote{{\tt gansukh@jejunu.ac.kr}}
and
Dong-han Yeom$^{b,d,e}$\footnote{{\tt innocent.yeom@gmail.com}}
}

\affiliation{
$^{a}$Department of Mathematics and Statistics, Masaryk University, 601 77 Brno, Czech Republic\\
$^{b}$Department of Physics Education, Pusan National University, Busan 46241, Republic of Korea\\
$^{c}$Department of Science Education, Jeju National University, Jeju 63243, Republic of Korea\\
$^{d}$Department of Physics and Astronomy, University of Waterloo, Waterloo, ON N2L 3G1, Canada\\
$^{e}$Leung Center for Cosmology and Particle Astrophysics, National Taiwan University, Taipei 10617, Taiwan
}

\begin{abstract}
We construct a canonical quantization, the Wheeler-DeWitt equation, of the interior geometry of static and spherically symmetric black holes in Einstein–Maxwell–$\Lambda$ framework, focusing on Reissner–Nordstr\"{o}m. The wave function of the Wheeler-DeWitt equation for the Reissner-Nordstr\"{o}m black hole is set to be on-shell and exhibiting exponential damping away from the classical locus. Horizon boundary conditions for the wave function generate two regimes: a single inward mode from event horizon yields monotonic decay, while superpositions produce either a quantum bounce (single time arrow) or interference‑driven “annihilation‑to‑nothing” (two time arrows). We show that these are generic features of static black hole interiors. Furthermore, the wave function of the Schwarzschild black hole, obtained as the charge-neutral limit of the Reissner–Nordstr\"{o}m black hole, is monotonically decaying and no longer unbounded. Moreover, this framework unifies classical and quantum interiors, suggests a quantum gravitational suppression to the mass inflation, and motivates extensions to Kerr and regular black holes.
\end{abstract}

\maketitle

\newpage

\tableofcontents

\section{Introduction\label{sec:intro}}
The interior regions of black holes provide a natural laboratory in which the principles of general relativity are pushed to their limits, indicating the need for a quantum theory of gravity. The classical prediction of a spacetime singularity~\cite{Hawking:1970zqf}, where the classical evolution breaks down~\cite{Hawking:1976ra}, is widely expected to be resolved by quantum effects~\cite{Boehmer:2007ket,Modesto:2006mx,Zhang:2023yps}. Clarifying the mechanism of this resolution remains a central challenge in theoretical physics, with implications for cosmic censorship and the ultimate fate of gravitational collapse~\cite{Chew:2023upu,Carballo-Rubio:2024dca,Cardoso:2017soq}. Building on the framework of quantum properties developed initially for Schwarzschild interiors~\cite{Bouhmadi-Lopez:2019kkt}, we extend the analysis to the broader charged family of Reissner–Nordstr\"{o}m black holes, which possess an additional inner (Cauchy) horizon and a richer causal structure but without singularities. Additional developments and related extensions are presented in Refs.~\cite{Brahma:2021xjy,Yeom:2021bpd,Chien:2023kqw,Singh:2024kcn,Kan:2022ism}. 

In this work, we focus on the interior dynamics of static and spherically symmetric black holes, specifically within the framework of Einstein-Maxwell-$\Lambda$ theory. Inside the event horizon, the roles of time and space coordinates are interchanged. For Reissner-Nordstr\"{o}m black holes, the interior dynamics are tightly linked to the instability of the inner Cauchy horizon, where the infinite blueshift of perturbations drives mass inflation and undermines both the Reissner–Nordstr\"{o}m interior and linear perturbation theory, necessitating fully nonlinear analyses~\cite{Poisson:1989zz,Burko:1997zy,McMaken:2023tft}. Following established approaches to the Schwarzschild interior, we extend the analysis to the charged case to explore how the electromagnetic field influences the interior quantum dynamics.

Our methodology involves a canonical quantization tailored to the black hole interior. We begin with the full Einstein-Maxwell-$\Lambda$ action, deriving a reduced Lagrangian from an anisotropic metric (the Kantowski-Sachs metric~\cite{Kantowski:1966te}) parametrized by scale factors $\{a(t), b(t)\}$ and a gauge field $Q(t)$ induced by the electromagnetic field. From this, we construct the Hamiltonian and obtain the corresponding Wheeler-DeWitt (WDW) equation. The Wheeler-DeWitt equation provides a conservative framework for describing the quantum evolution of the corresponding geometry~\cite{Cavaglia:1994yc,Garcia-Compean:2001jxk,Lopez-Dominguez:2006rou,Bastos:2007bg}. Analogous to the Schr\"{o}dinger equation, the present equation treats the configuration-space variables $\{X:=\ln a(t),Y:=\ln b(t),Q\}$ equally, with the $\{X,Y\}$-directions serving as spacelike variables and the $Q$- direction serving as a timelike variable for the interior evolution. Other related works on quantizing charged black holes using the Wheeler–DeWitt equation can be found in Refs.~\cite {Moniz:1997ur,Garattini:2008qs,Blacker:2023ezy}.

Specializing to the Reissner-Nordstr\"{o}m interior, we solve the Wheeler-DeWitt equation and construct wave functions localized on the classical trajectories connecting the event horizon $r_+$ and the Cauchy horizon $r_-$. We investigate two primary physical scenarios arising from different boundary conditions at the horizons:
\begin{enumerate}
    \item [(i)] A single inward-propagating component from $r_+$, corresponding to a single arrow of time and allowing for monotonic decay (although the wave function for the Schwarzschild black hole case is unbounded, we propose that the wave function for a charge-neutral Reissner–Nordstr\"{o}m black hole is bounded).
    \item [(ii)] A superposition of components from both horizons. If a single arrow of time is maintained, the solution exhibits a quantum bounce; if counter-propagating components are present, their interference can lead to an “annihilating-to-nothing”~\cite{Bouhmadi-Lopez:2019kkt}. 
\end{enumerate}
By analyzing the structure of the wave function, we demonstrate that the Reissner-Nordstr\"{o}m interior exhibits a monotonic decay, a quantum bounce, or an “annihilating-to-nothing” localized at a timelike $Q$ coordinate, analogous to the behavior found in the simpler Schwarzschild case. This suggests that these are generic features of static black hole interiors~\cite{Bouhmadi-Lopez:2019kkt,Brahma:2021xjy,Yeom:2021bpd,Chien:2023kqw}, see Table~{\ref{tab:BHI}}. 

\begin{table}[h]
    \centering
    \begin{tabular}{|c|c|c|c|}
        \hline
         &  Monotonic Decay & Quantum Bounce &  Annihilation-to-Nothing\\ \hline
         Schwarzschild BH&  \includegraphics[width=0.2\textwidth,keepaspectratio]{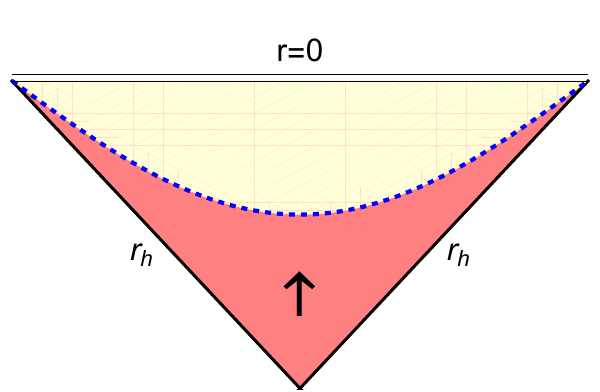}&  \includegraphics[width=0.2\linewidth,keepaspectratio]{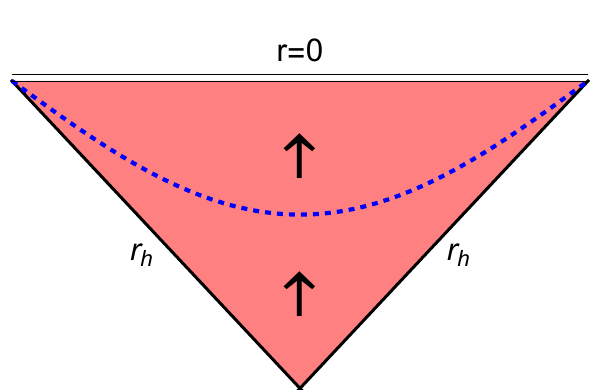}& \includegraphics[width=0.2\linewidth,keepaspectratio]{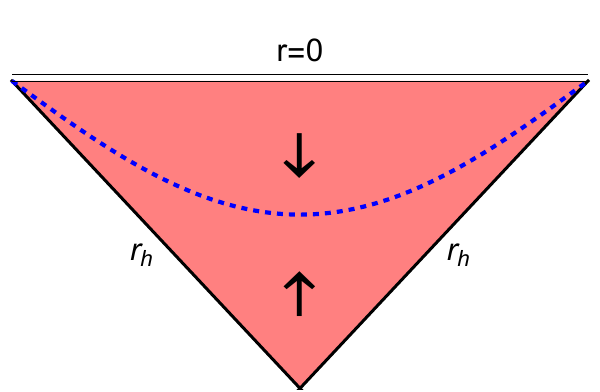}\\ \hline
         Schwarzschild-(A)dS BH& \includegraphics[width=0.2\linewidth,keepaspectratio]{SPD1.pdf}& \includegraphics[width=0.2\linewidth,keepaspectratio]{SPD2.pdf}& \includegraphics[width=0.2\linewidth,keepaspectratio]{SPD3.pdf}\\ \hline
        Reissner--Nordstr\"{o}m BH& \includegraphics[width=0.2\linewidth,keepaspectratio]{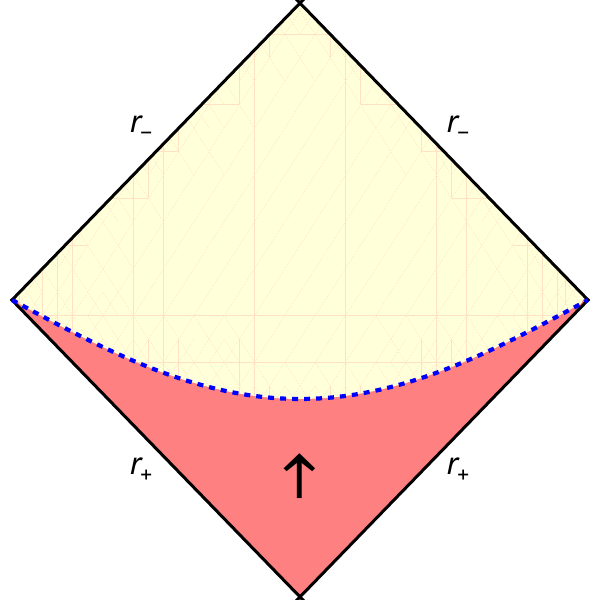}& \includegraphics[width=0.2\linewidth,keepaspectratio]{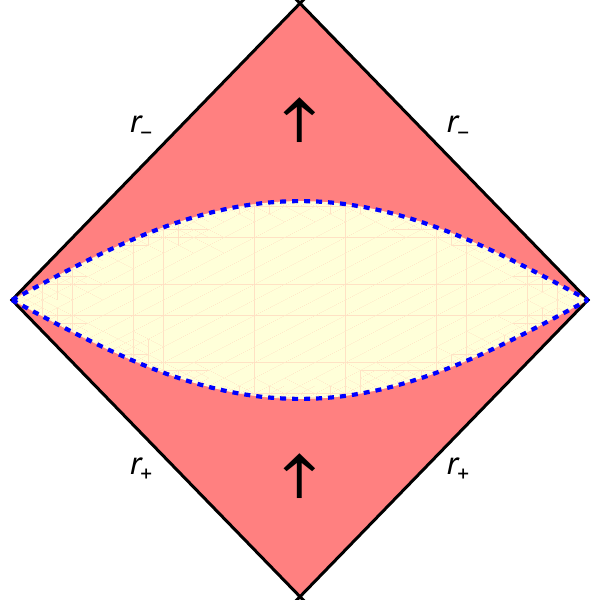}& \includegraphics[width=0.2\linewidth,keepaspectratio]{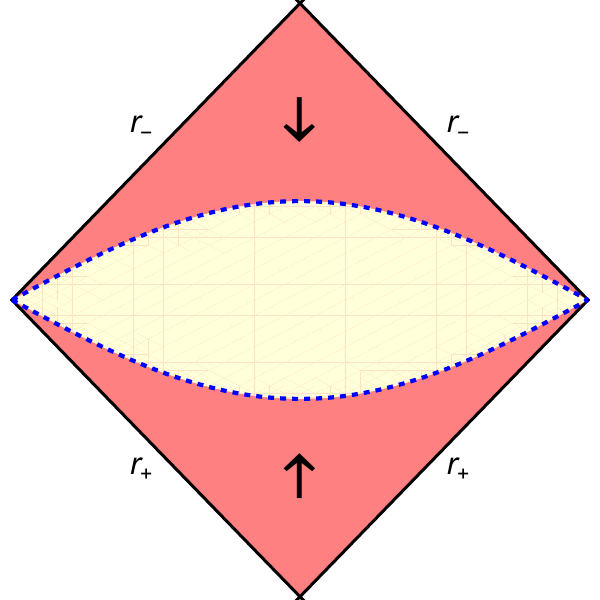}\\ \hline
    \end{tabular}
    \caption{Possible Penrose diagrams of black hole interiors are displayed. The arrows indicate the direction of time; the red region denotes spacetime containing waves, while the yellow region denotes wave-free (empty) spacetime.}
    \label{tab:BHI}
\end{table}

In relation to the Cauchy horizon and the phenomenon of mass inflation, a complementary quantum-gravitational suppression is proposed: (i) Interpret the interior wave function as imposing an “annihilation-to-nothing” boundary condition, so that both the event and Cauchy horizons lie in the causal past of the bounce. In this picture, the infinite blueshift at the would-be Cauchy horizon is not operationally relevant. (ii) With only an ingoing mode from the event horizon, no physical structure forms at the Cauchy horizon. While not a classical completion, an appropriate boundary condition on the wave function smoothly ameliorates the interior geometry and removes the mechanism for mass inflation. Whether this mechanism extends to rotating or regular black holes with inner horizons (e.g., Kerr or regular charged solutions) remains open, and we leave a systematic treatment of these cases to future work.

This paper is organized as follows. In Sec.~\ref{sec:wdw}, we present the Einstein-Maxwell-$\Lambda$ framework and the Wheeler-DeWitt equation. In Sec.~\ref{sec:RNBH}, we specialize to the Reissner-Nordstr\"{o}m interior and analyze the associated classical trajectories. In Sec.~\ref{sec:WFS}, we construct bounded quantum solutions and discuss boundary conditions at the horizons.  In Sec.~\ref{sec:AT}, we analyze the physical interpretation in terms of the arrow of time. In Sec.~\ref{sec:CH}, we comment on the Cauchy horizon and mass inflation within this framework. Finally, in Sec.~\ref{sec:con} we conclude with a summary and an outlook on the quantum nature of black-hole interiors.

\section{Wheeler-DeWitt Equation for Static Black Hole Interiors\label{sec:wdw}}
We investigate the interior geometry of black holes, focusing on the spacetime region inside the event horizon where the roles of time-like and space-like coordinates interchange. Our analysis is confined to such a region. Throughout, we adopt natural units with $c = G = \hbar = 1$, and the metric signature is taken to be $(-,+,+,+)$.

The dynamics are governed by the Einstein-Maxwell theory on a four-dimensional Lorentzian manifold $(M,g_{\mu\nu})$ minimally coupled to an Abelian gauge field $A_\mu$ with field strength $F_{\mu\nu}$. The action is
\begin{eqnarray} \label{eq:S}
    S=\frac{1}{16\pi}\int d^4x\sqrt{-g}(R-F^{\mu\nu}F_{\mu\nu}-2\Lambda),
\end{eqnarray}
where $R$ is Ricci scalar of $g_{\mu\nu}$ and $\Lambda$ is the cosmological constant. 
Varying the action with respect to $g_{\mu\nu}$ and $A_\mu$ yields the Einstein and Maxwell equations,
\begin{eqnarray}\label{eq:eom1}
    G_{\mu\nu}+\Lambda g_{\mu\nu}=8\pi T_{\mu\nu},\quad \nabla_\mu F^{\mu\nu}=0,
\end{eqnarray}
with the Einstein tensor and the electromagnetic stress-energy tensor
\begin{eqnarray}\label{eq:eom2}
    G_{\mu\nu}&=&R_{\mu\nu}-\frac{1}{2}g_{\mu\nu}R,\\
    T_{\mu\nu}&=&\frac{1}{4\pi}\left(g^{\alpha\beta}F_{\mu\alpha}F_{\nu\beta}-\frac{1}{4}g_{\mu\nu}F_{\alpha\beta}F^{\alpha\beta}\right).
\end{eqnarray}

Inside a static black hole, we choose coordinates adapted to the interior causal structure and take the electromagnetic potential and field strength as
\begin{eqnarray} \label{eq:FA}
    F_{\mu\nu}=\partial_\mu A_\nu-\partial_\nu A_\mu,\quad A_\mu=\left(0,\frac{r_Q}{t},0,0\right),
\end{eqnarray}
where $r_Q^2={q^2}/{4\pi\epsilon_0}$ is a characteristic length scale with respect to charge $q$. 

The metric ansatz of the spacetime inside a static black hole takes the form
\begin{eqnarray}\label{eq:sbh}
    ds^2= -\left(\frac{\Lambda}{3}t^2+\frac{r_s}{t}-Q^2 (t)-1\right)^{-1} dt^2+\left(\frac{\Lambda}{3}t^2+\frac{r_s}{t}-Q^2 (t)-1\right) dR^2+t^2d\Omega^2,
\end{eqnarray}
where $r_s=2M$ is Schwarzschild radius, and the on-shell classical solution has $Q(t)={r_Q}/{t}$. Special cases are recovered as follows: Schwarzschild for $Q=\Lambda=0$, Schwarzschild-(anti-) de Sitter for $Q=0$, and Reissner–Nordstr\"{o}m for $\Lambda=0$. We will allow $Q(t)$ to fluctuate off-shell to accommodate quantum effects.

A diffeomorphism maps the above interior metric to an anisotropic form~\cite{Kantowski:1966te},
\begin{eqnarray}\label{eq:ani}
    ds^2= -{N^2}(t)dt^2+{a^2}(t)dR^2+r_s^2 \frac{{b^2}(t)}{{a^2}(t)}d\Omega_2^2,
\end{eqnarray}
where $N(t)$ is the lapse function and $\{a(t),b(t)\}$ are positive, dimensionless scale factors. Defining $X:=\ln{a(t)}$ and $Y:=\ln{b(t)}$, matching to the interior solution imposes the algebraic constraint
\begin{eqnarray}\label{eq:XYQL}
    e^X+e^{-X}=e^{-Y}-Q^2 e^{-X}+\frac{\Lambda}{3}r_s^2 e^{2Y-3X}.
\end{eqnarray}
This relation defines a constraint surface in $\{X,Y,Q\}$ on which classical interior trajectories lie.

Substituting the anisotropic ansatz into the action and integrating by parts yields the reduced Lagrangian
\begin{eqnarray} \label{eq:L}
    \mathcal{L}=\frac{r_s^2b^2}{Na}\left[\frac{N^2a^2}{r_s^2b^2}+\frac{\dot{a}^2}{a^2}-\frac{\dot{b}^2}{b^2}+\frac{\dot{Q}^2}{a^2}-N^2\Lambda\right].
\end{eqnarray}
Introducing canonical momenta $p_{x_i}=\partial \mathcal{L}/\dot{x_i}$ for $x_i=\{a,b,Q,N\}$, the Hamiltonian is 
\begin{eqnarray} \label{eq:H}
    \mathcal{H}=\frac{Na}{4r_s^2b^2}\left[a^2p_a^2-b^2p_b^2+a^2p_Q^2-4r_s^2b^2+4r_s^4 \Lambda\frac{b^4}{a^2}\right]+\lambda_N\frac{p_N}{r_s^2},
\end{eqnarray}
where $\lambda_N$ is a Lagrange multiplier of the constraint $p_N$.

Quantizing by $p_{x_i}\to\hat p_{x_i}=-i\partial_{x_i}$ and switching to logarithmic variables $\{X,Y\}$, the Wheeler–DeWitt equation governing the interior wave function $\Psi(X,Y,Q)$ is further derived as
\begin{eqnarray} \label{eq:wdw}
    \left[\frac{\partial^2}{\partial X^2}-\frac{\partial^2}{\partial Y^2}+e^{2X}\frac{\partial^2}{\partial Q^2}+4r_s^2e^{2Y}-4 r_s^4\Lambda e^{4Y-2X}\right] \Psi(X,Y,Q)=0.
\end{eqnarray}
In the $\{Q,\Lambda\}\to0$ limit, the reduced dynamics reproduces the Schwarzschild interior as shown in~\cite{Bouhmadi-Lopez:2019kkt}; for $Q\to0$, it gives the Schwarzschild-(anti-) de Sitter interior and its spacetime beyond the cosmological horizon as shown in~\cite{Chien:2023kqw}; for $\Lambda=0$, it yields the Reissner–Nordstr\"{o}m interior. We investigate the quantum properties of the wave function $\Psi(X,Y,Q)$ of the Reissner–Nordstr\"{o}m interior in the following section.

\section{Constraint Surface and Horizon Asymptotics for Reissner-Nordstr\"{o}m Interiors\label{sec:RNBH}}
In this section, we specialize to Reissner–Nordstr\"{o}m black holes by setting $\Lambda=0$ in Sec.~\ref{sec:wdw} and restricting attention to the interior region $r_-\leq t\leq r_+$. The two horizons are located at
\begin{eqnarray} \label{eq:hrz}
    r_{\pm}=\frac{1}{2}\left(r_s\pm\sqrt{r_s^2-4r_Q^2}\right),
\end{eqnarray}
with $r_+$ the event horizon and $r_-$ the Cauchy horizon. Throughout, we assume $2r_Q<r_s$ so that both horizons exist and the interior region is well-defined for our analysis of quantum properties.

The constraint sphere Eq.~(\ref{eq:XYQL}) for Reissner–Nordstr\"{o}m black holes simplifies to
\begin{eqnarray}\label{eq:XY}
e^X+e^{-X}=e^{-Y}-Q^2 e^{-X}.
\end{eqnarray}
This relation defines the surface on which classical interior trajectories lie.

Classical trajectories parameterized by the coordinate $t$ and labeled by the charge scale $r_Q$ admit the parametric form
\begin{eqnarray}\label{eq:pl}
    \{X,Y,Q\}=\left\{ \log \left(\sqrt{\frac{{r_s}}{t}-\frac{{r_Q}^2}{t^2}-1}\right),\log \left(\frac{t}{r_s}\sqrt{\frac{{r_s}}{t}-\frac{{r_Q}^2}{t^2}-1}\right),\frac{r_Q}{t} \right\},\quad r_-\leq t\leq r_+.
\end{eqnarray}
These trajectories encode the interior evolution between the two horizons on the constraint surface.

The parametric curves around $r_\pm$ approach straight asymptotes given by
\begin{eqnarray} \label{eq:al}
    X-Y=-\log\left(\frac{r_\pm}{r_s}\right),\quad Q=\frac{r_Q}{r_\pm}.
\end{eqnarray}
Thus, each classical trajectory intersects the corresponding asymptotic line with a charge-dependent value $Q=r_Q/r_\pm$, while sharing the same slope $X-Y$ set by $r_\pm/r_s$.

\begin{figure}[!h]
    \centering
    \includegraphics[width=0.45\textwidth]{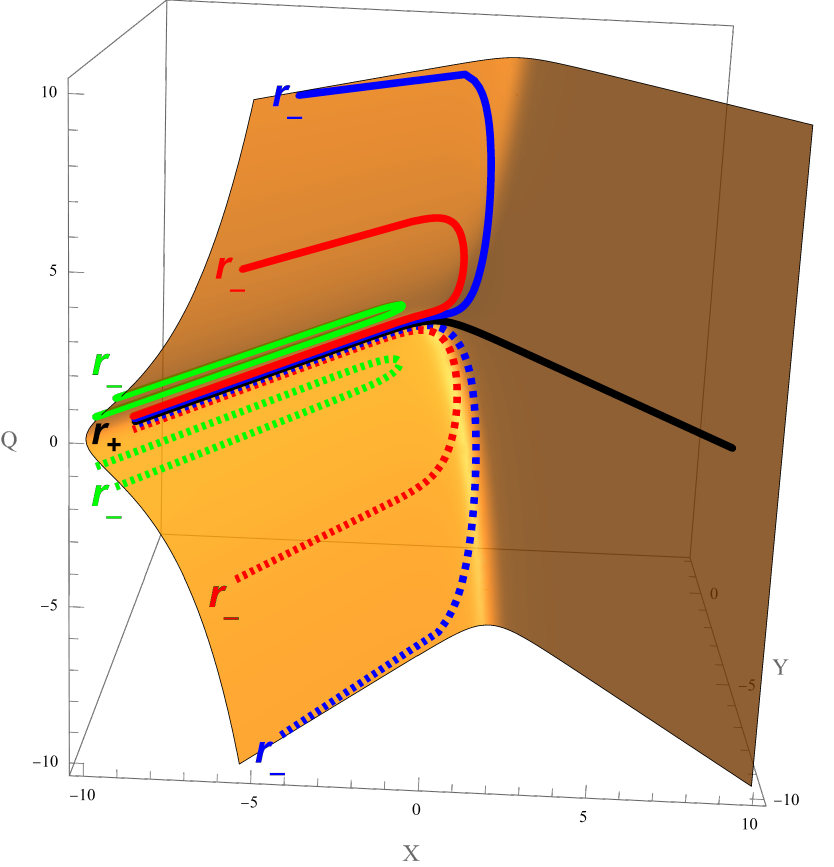}
    \caption{The yellow surface is the constraint surface Eq.~(\ref{eq:XY}) for the Reissner–Nordstr\"{o}m black hole. The colored lines are Eq.~(\ref{eq:pl}) with $r_Q=\pm0.48$(green), $r_Q=\pm0.2$(red) and $\pm0.1$(blue), where the solid lines are $r_Q$ with $+$ and dotted lines are $-$. The black line is the Schwarzschild black hole, $r_Q=0$.} 
    \label{fig:CT}
\end{figure}

In the regime $r_Q\ll r_s$, one has $\log(r_+/r_s)\simeq0$, whereas $\log(r_-/r_s)$ retains a pronounced dependence on $r_Q$. Consequently:
\begin{itemize}
    \item Ingoing trajectories emerging from the event horizon $r_+$ bunch near the line $X=Y$ with $Q\simeq0$, indicating that waves entering from $r_+$ are focused toward the vicinity of this line.
    \item Outgoing trajectories toward the Cauchy horizon $r_-$ separate according to $r_Q$ via $Q=r_Q/r_-$, leading to a charge-dependent spread as the evolution proceeds toward $r_-$.
\end{itemize}
One can see that classical curves with different $r_Q$ converge around $X=Y$, $Q\simeq0$, while near $r_-$ they fan out with distinct $Q$-asymptotes determined by $r_Q/r_-$, see Fig.~\ref{fig:CT}.

\section{Analytic Solution of Wheeler-DeWitt Equation for Reissner-Nordstr\"{o}m Interiors\label{sec:WFS}}
We adopt the semiclassical prescription that physically relevant Wheeler–DeWitt states are sharply peaked on classical solutions, in the sense that the maximal amplitude of wavefunction follows the classical trajectory on the constraint surface. Operationally, we refer to this locus of the maximal amplitude as the steepest-descent line of the wavefunction amplitude. In the interior of the Reissner–Nordstr\"{o}m black holes, the steepest-descent condition is implemented as an on-shell requirement: the wavefunction is maximally supported on the classical trajectory Eq. (\ref{eq:pl}) between the horizons $r_-\leq t\leq r_+$.

\subsection{General Solution\label{sec:GS}}
Such a solution to the Wheeler-Dewitt equation for Reissner–Nordstr\"{o}m Black Holes (Eq.~(\ref{eq:wdw}) with $\Lambda=0$) can be constructed by separation of variables, $\Psi(X,Y,Q)=\phi(X)\psi(Y)\varphi(Q)$. The separation of variables for a certain choice of separation constants $\lambda,k\in [0,\infty)$ follows
\begin{eqnarray} \label{eq:SVEQ}
    \frac{d^2\phi}{d X^2}- 4\lambda^2 e^{2X}\phi+k^2\phi &=&0,\\
    \frac{d^2\psi}{d Y^2}-4r_s^2e^{2Y}\psi+k^2\psi&=&0,\\
    \frac{d^2\varphi}{d Q^2}+4\lambda^2\varphi&=&0.
\end{eqnarray}
The differential equations can be solved individually as
\begin{eqnarray} \label{eq:sol}
\phi(X)&=&c_1 I_{ik}(2\lambda e^X)+c_2 K_{ik}(2\lambda e^X),\\
\psi(Y)&=&c_3 I_{ik}(2r_s e^Y)+c_4 K_{ik}(2r_s e^Y),\\
\varphi(Q)&=&c_5e^{+ 2i\lambda Q}+c_6e^{- 2i\lambda Q}.
\end{eqnarray}
where $I_{ik}$ and $K_{ik}$ are the modified Bessel function. 
To impose normalizability at a large argument, we set the coefficients of the growing solutions to zero: since the modified Bessel function $I_{ik}(z)$ diverges as $z\to\infty$, we take $c_1=c_3=0$ so that the wave function remains bounded. Using the asymptotic behavior $K_{ik}(z)\sim \sqrt{\pi/(2z)}e^{-z}$ for $z\to\infty$, the surviving components are exponentially damped in the regions $X>0$ and $Y>0$, respectively. Consequently, the only non-negligible contributions arise from waves sourced at the horizons $r_\pm$ and guided along the classical trajectory Eq.~(\ref{eq:pl}).

The general solution of the bounded wave function can be written as
\begin{eqnarray} 
    \Psi(X,Y,Q) &=&\int^\infty_{0}\int^\infty_{0} \left(f_1(k,\lambda)e^{2i\lambda Q }+f_2(k,\lambda)e^{-2i\lambda Q }\right)K_{ik}(2\lambda e^X)K_{ik}(2r_s e^Y)d\lambda dk,\\
    &=&\int^\infty_{0}\int^0_{-\infty} f_1(k,-\lambda)e^{-2i\lambda Q }K_{ik}(-2\lambda e^X)K_{ik}(2r_s e^Y) d\lambda dk \nonumber\\
    &&+\int^\infty_{0}\int^\infty_{0} f_2(k,\lambda)e^{-2i\lambda Q }K_{ik}(2\lambda e^X)K_{ik}(2r_s e^Y) d\lambda dk,\\
    &=&\int^\infty_{0}\int^\infty_{-\infty} f(k,\lambda)e^{-2i\lambda Q }K_{ik}(2\lvert\lambda\lvert e^X)K_{ik}(2r_s e^Y)d\lambda dk,\label{eq:gsb}
\end{eqnarray}
where $f(k,\lambda):= f_2(k,\lambda)=f_1(k,-\lambda)$ and the property of absolute value are used. In principle, various functional forms may be chosen for $f(k)$. A frequently employed choice is a Gaussian wave packet, but alternative options can also be considered.

The interpretation of the general bounded wave function for the Reissner–Nordstr\"{o}m interior, Eq.~(\ref{eq:gsb}), is a natural extension of the Schwarzschild case discussed in Ref.~\cite{Bouhmadi-Lopez:2019kkt}. In this formulation, the state $\Psi(X,Y,Q)$ obeys a Schr\"{o}dinger-type equation on the three-dimensional configuration space $\{X,Y,Q\}$, where separation of variables highlights a clean split between a “time-like” section and a “spatial” section. In particular, the factor $\varphi(Q)$ plays the role of a time-like evolution component along the $Q$-direction, while the remaining product $\phi(X) \psi(Y)$ satisfies a time-independent Schr\"{o}dinger equation on the two-dimensional subspace spanned by $\{X,Y\}$. The effective potentials along the $X-$ and $Y-$directions are governed by the barrier terms $e^{2X}$ and $e^{2Y}$, respectively, which control the localization and exponential damping structure of the bounded solutions in each direction.

\subsection{Gaussian Wave Packet\label{sec:EGS}}
The general solution in Eq.~(\ref{eq:gsb}) involves double integrals, which complicate the analysis. To streamline the analysis, we invoke a functional identity and take a Gaussian wave packet to enable the desired reduction.

We assume the identity~\cite{gradshteyn2014table}
\begin{equation}\label{eq:ide}
    \int ^{\infty}_0{x \tanh (\pi  x) K_{i x}(a) K_{i x}(b)}{dx}=\frac{\pi}{2} \sqrt{a b} \frac{e^{-(a+b)}}{a+b},
\end{equation}
and 
\begin{equation}\label{eq:fkl}
    f(k,\lambda)={k \tanh (\pi k)}g(\lambda),
\end{equation}
so that it is reduced to
\begin{equation}\label{eq:gsbs}
    \Psi(X,Y,Q)=\int^\infty_{-\infty} g(\lambda)e^{-2i\lambda Q} \left(\frac{\pi}{2}\sqrt{(2| \lambda | e^X )(2r_s e^Y)}\frac{e^{-2(| \lambda | e^X+r_s e^Y)}}{2| \lambda | e^X+2r_s e^Y}\right)d\lambda.
\end{equation} 
We may choose $g(\lambda)$ to be a Gaussian wave packet as
\begin{equation}\label{eq:gw}
    g(\lambda)=Ae^{-\frac{1}{2}\sigma^2(\lambda-|\lambda_0|)^2} e^{2i\lambda Q_0},
\end{equation}
where $A$ denotes the normalization constant, $\sigma$ is the standard deviation characterizing the width of the packet, and $\{\lambda_0,Q_0\}$ determine the center of localization in the chosen variables. Notice that this is not an exclusive choice, as other forms of $f(k)$ may work equally well.

To understand the structure of the wave function, we begin by decomposing it into its constituent components. As a first step, we normalize the wave function along the asymptotic line given by Eq.~(\ref{eq:al}). Explicitly, we impose
\begin{equation}\label{eq:lmt}
    \lim_{X\to-\infty}\left|\Psi\left(X,X+\log(\frac{r_{\pm}}{r_s}),\frac{r_Q}{r_{\pm}}\right)\right|^2=1\quad \Rightarrow \quad A=\frac{2\sqrt{2}\sigma}{\pi^{3/2}}.
\end{equation}
Second, we focus on the bracketed term by fixing $\lambda$ at an arbitrary reference value $\lambda_0$ in Eq.~(\ref{eq:gsbs}), 
\begin{equation}\label{eq:kk}
    \frac{\pi}{2}\sqrt{(| \lambda_0 | e^X )(r_s e^Y)}\frac{e^{-2(| \lambda_0 | e^X+r_s e^Y)}}{| \lambda_0 | e^X+r_s e^Y}.
\end{equation} 
It is centered at $X-Y=-\log\left(\frac{|\lambda_0|}{r_s}\right)$. Comparison with the asymptotic relations in Eq.~(\ref{eq:al}) shows that $|\lambda_0|=r_\pm$ and $Q_0=r_Q/r_\pm(=: Q_{0\pm})$. As a third step, we temporarily disregard the $\lambda$-dependent terms in the brackets to isolate and examine the intrinsic Gaussian behavior. The Fourier transform of a Gaussian amplitude yields another Gaussian in $Q$-space as
\begin{equation}\label{eq:gwq}
    \int^\infty_{-\infty}  e^{-\frac{1}{2}\sigma^2(\lambda-|\lambda_0|)^2} e^{-2i\lambda (Q-Q_0)}d\lambda=\sqrt{\frac{2 \pi }{\sigma ^2}}e^{-\frac{2 (Q-{Q_0})^2}{\sigma ^2}}e^{-2 i |{\lambda_0}| (Q-Q_0)}.
\end{equation}
The Gaussian envelope falls off exponentially away from $Q_0$, indicating strong localization in the $Q$-direction. While this simple form cannot be applied to the complete solution in Eq.~(\ref{eq:gsbs}), we anticipate that the exact wave function exhibits the same characteristic suppression at large $|Q-Q_0|$.

In conclusion, the simplified Gaussian wave function for the Reissner–Nordstr\"{o}m interior constitutes a particular solution to the Wheeler–DeWitt equation, Eq.~(\ref{eq:wdw}), in the absence of a cosmological constant. Its functional form can be written as
\begin{equation}\label{eq:gbwf}
    \Psi_\pm(X,Y,Q)=\sqrt{\frac{2}{\pi}}\sigma\int^\infty_{-\infty}  e^{-\frac{1}{2}\sigma^2(\lambda-r_\pm)^2} e^{-2i\lambda (Q-Q_{0\pm})} \sqrt{(| \lambda | e^X )(r_s e^Y)}\frac{e^{-2(| \lambda | e^X+r_s e^Y)}}{| \lambda | e^X+r_s e^Y}d\lambda.
\end{equation} 
where, naggingly repeating, $\sigma$ is the standard deviation, $Q_{0\pm}=r_Q/r_\pm$ are the localization centers in the $Q$–direction, $r_Q$ is the characteristic length scale associated with the charge, $r_\pm$ are the outer (event) and inner (Cauchy) horizons of the Reissner–Nordstr\"{o}m black hole, $r_s$ is the Schwarzschild radius, and $\lambda$ is a separation constant. For the incoming wave originating from the event horizon, we denote the solution by $\Psi_{+}$, characterized by $r_+$ and $Q_{0+} = r_Q / r_+$. For the outgoing(ingoing) wave propagating toward(from) the Cauchy horizon, we denote the solution by $\Psi_{-}$, characterized by $r_-$ and $Q_{0-} = r_Q / r_-$.

\subsection{Wave Function and Its Boundary Conditions\label{sec:EGSe}}
\begin{figure}[h]
    \begin{center}
    \includegraphics[width=0.45\textwidth]{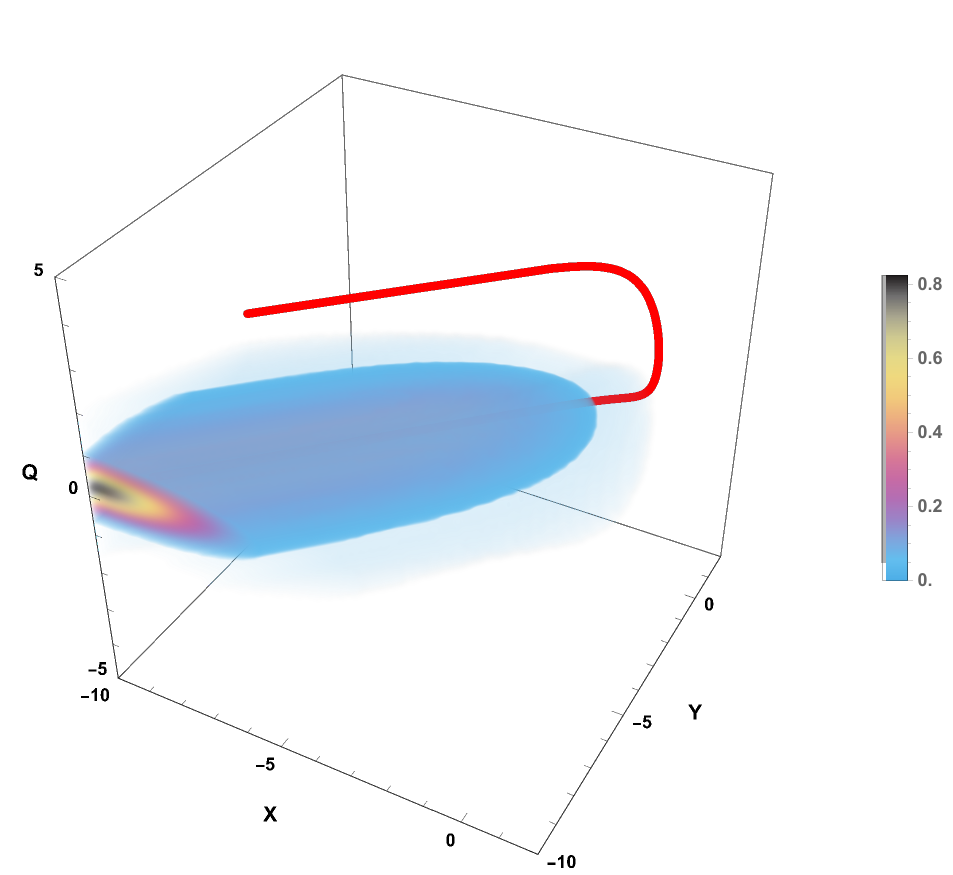}\hspace{0.05 cm}
    \includegraphics[width=0.45\textwidth]{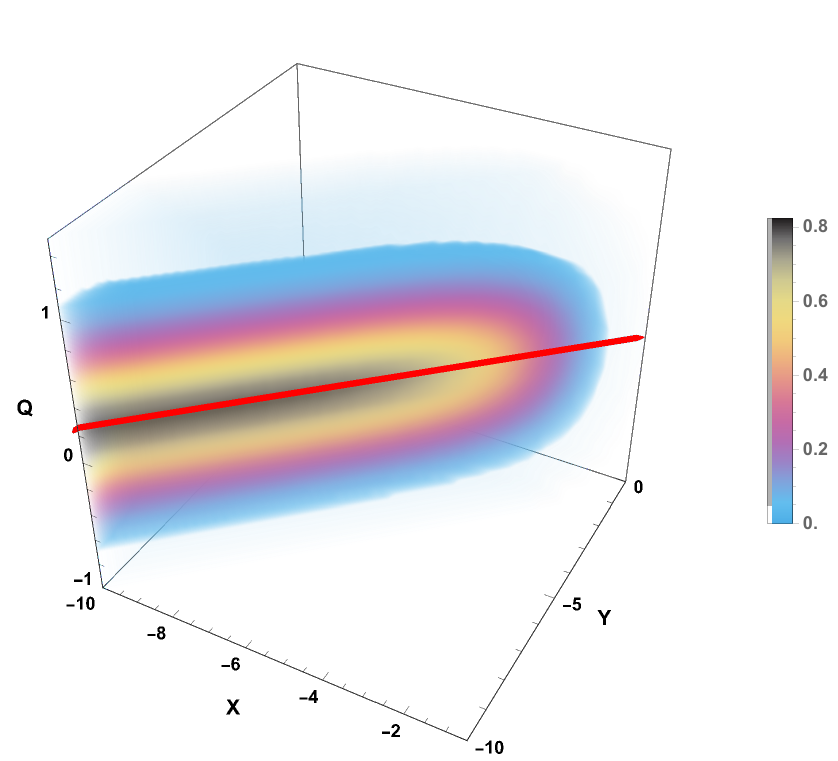}\\    
    \caption{Left: The squared modulus of the wave function $\Psi_+$ Eq.(\ref{eq:gbwf}) with $\{r_s,\sigma,r_Q\}=\{1,1,0.2\}$ is shown. The red line is the classical trajectory. Right: The planar cross-section shows that the amplitude maximum traces the steepest-descent contour.} 
    \label{fig:WF1}
    \end{center}
\end{figure}

\begin{figure}[h]
    \begin{center}
    \includegraphics[width=0.45\textwidth]{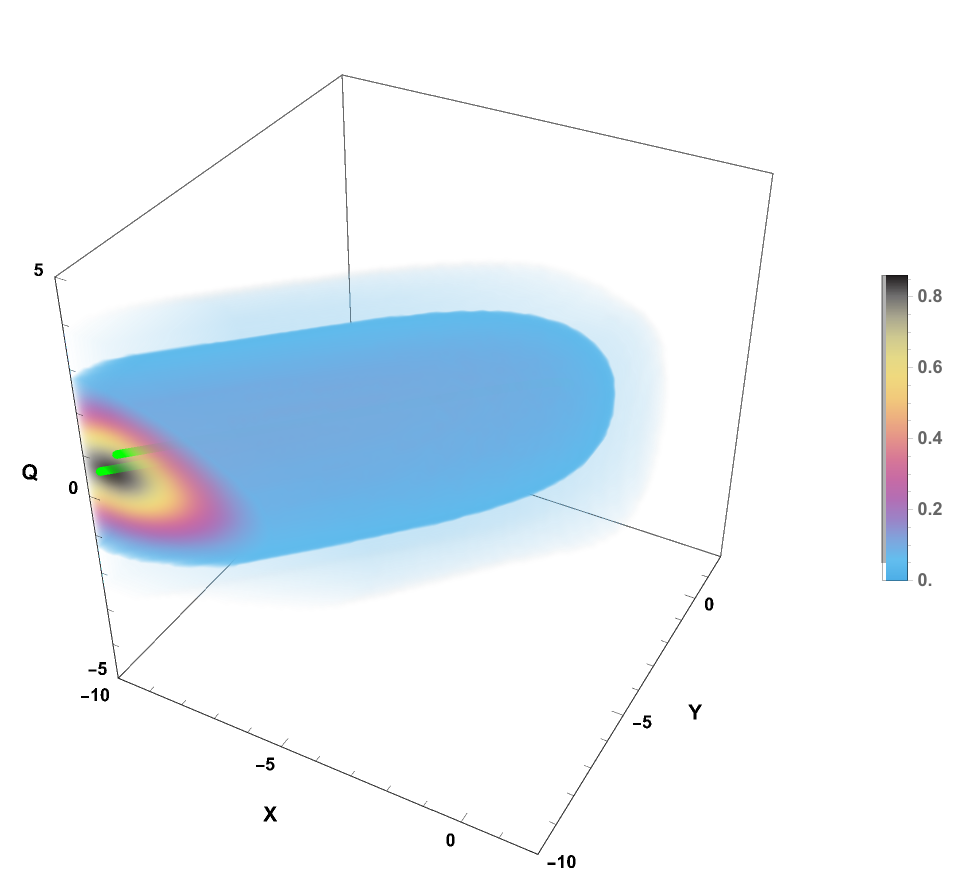}\hspace{0.05 cm}
    \includegraphics[width=0.45\textwidth]{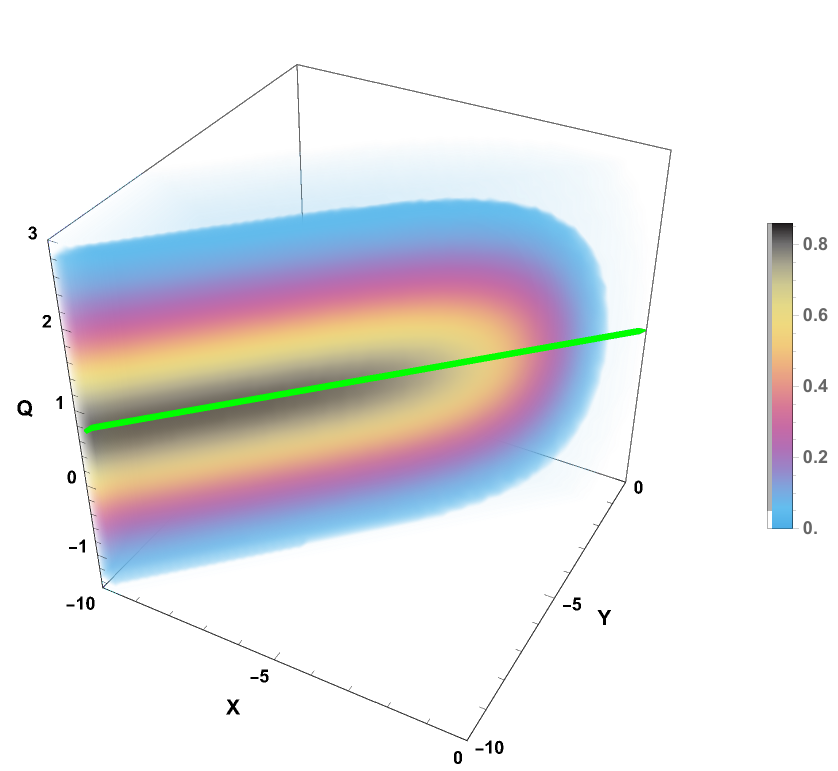}\hspace{0.05 cm}
    \includegraphics[width=0.45\textwidth]{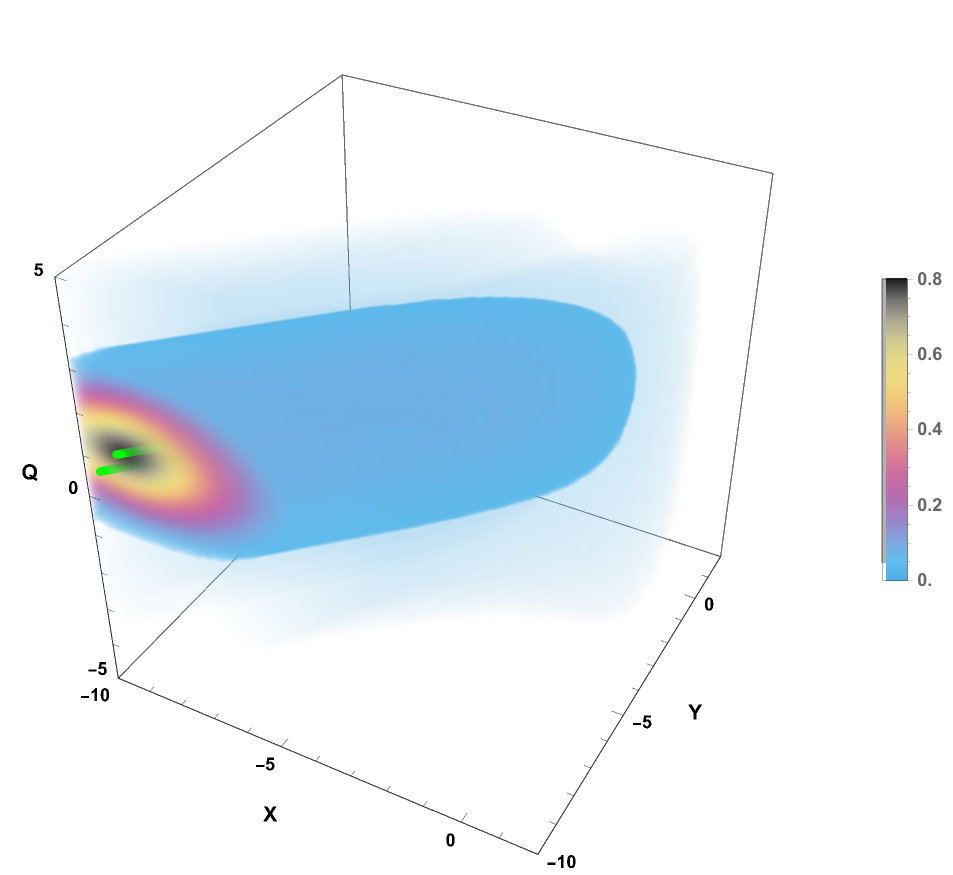}\hspace{0.05 cm}
    \includegraphics[width=0.45\textwidth]{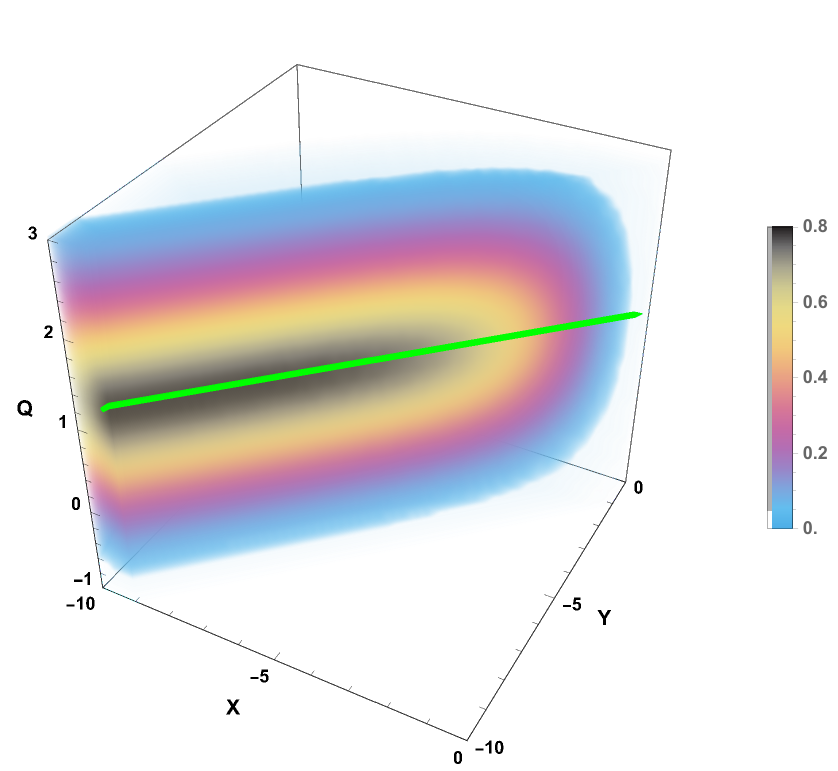}
    \caption{The squared modulus of the wave function Eq.(\ref{eq:gbwf}) with $\{r_s,\sigma,r_Q\}=\{1,2.3,0.48\}$ is shown. The green line is the classical trajectory. $\Psi_+$ corresponds to the upper panel and $\Psi_-$ to the lower; the right-hand inset shows a planar cross-section in which the amplitude maximum traces the steepest-descent contour.} 
    \label{fig:WF2}
    \end{center}
\end{figure}

From the requirement of continuity at the classical boundary $r_+$, the wave function $\Psi_+$ is expected to propagate inward from the event horizon. In contrast, there is no definitive prescription for the presence or absence of the inner-horizon component $\Psi_-$. In light of this uncertainty, we analyze two scenarios: the first involving only the inward-propagating component $\Psi_1=\Psi_+$ (see Fig.~\ref{fig:WF1}), and the second incorporating both components, $\Psi_2=\Psi_+ + \Psi_-$ (see Fig.~\ref{fig:WF2}).

We note that the choice of the width parameter $\sigma$ for the wave function $\Psi_{-}$ presents a numerical challenge. A small value of $\sigma$ yields a sharply localized Gaussian but can lead to numerical instability, while a large $\sigma$ improves numerical stability at the cost of producing an undesirably broad wave packet. In the regime $r_Q\ll r_s$, the inner-horizon radius $r_-$ becomes extremely small, so that the Gaussian factor $\exp(\sigma^2(\lambda-r_-)^2 /2)$ is effectively displaced from the intended location. To mitigate this, we adopt a larger $\sigma$ together with a larger $r_Q$. This adjustment, however, reduces the decay rate of the wave amplitude in the $Q$–direction, as seen in Fig.~\ref{fig:WF2}, leading to partial overlap between the two horizon-sourced components. This overlap should not be interpreted as a genuine physical effect of the Reissner–Nordstr\"{o}m interior, but rather as a consequence of numerical limitations, see Appendix~\ref{appendix:ni}; in the ideal physical scenario, the two waves from $r_{\pm}$ would remain sharply localized and well-separated, as depicted in Fig.~\ref{fig:WF1}.

\subsection{Neutral Black Hole Limit\label{sec:NBH}}
In Einstein–Maxwell gravity, an initially charged Kerr–Newman (or Reissner–Nordstr\"{o}m) black hole generically evolves toward effective electrical neutrality via complementary classical infall and quantum discharge. In realistic plasmas, the Coulomb-enhanced capture of oppositely charged particles quickly neutralizes the net charge, as supported by trajectory statistics that reveal higher infall probabilities for particles of opposite signs when no confining magnetosphere is present. Quantum discharge proceeds through a grand‑canonical Hawking spectrum with electrochemical potential, biasing emission so that like‑signed charge escapes to infinity and reduces charge. Even as Hawking temperature $T_H\to 0$ near extremality, the near‑horizon field drives nonthermal Schwinger pair creation that separates charges, yielding a net outward current. Collectively, these mechanisms imply that the charged black holes will be neutralized on large astrophysical timescales Refs.~\cite{Gibbons:1975kk,Page:1977um,Iso:2006wa,Gong:2019aqa}.

Within a Wheeler–DeWitt formulation, the Gaussian wave function describing the charge-neutral end-state of a Reissner–Nordstr\"{o}m black hole—i.e., the Schwarzschild limit—can be represented as
\begin{equation}\label{eq:gbwfe}
    \Psi_e(X,Y,Q)=\sqrt{\frac{2}{\pi}}\sigma\int^\infty_{-\infty}  e^{-\frac{1}{2}\sigma^2(\lambda-1)^2} e^{-2i\lambda Q} \sqrt{(| \lambda | e^X )(r_s e^Y)}\frac{e^{-2(| \lambda | e^X+r_s e^Y)}}{| \lambda | e^X+r_s e^Y}d\lambda.
\end{equation} 

\begin{figure}[h]
    \begin{center}
    \includegraphics[width=0.45\textwidth]{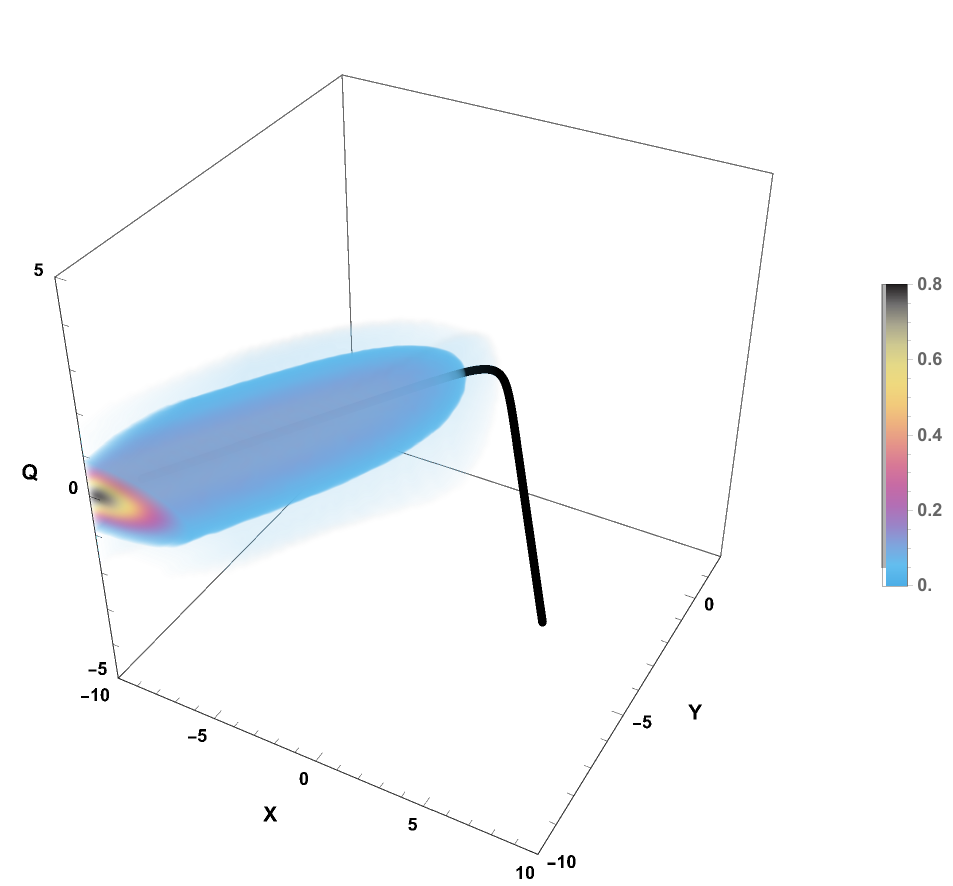}\hspace{0.05 cm}
    \includegraphics[width=0.45\textwidth]{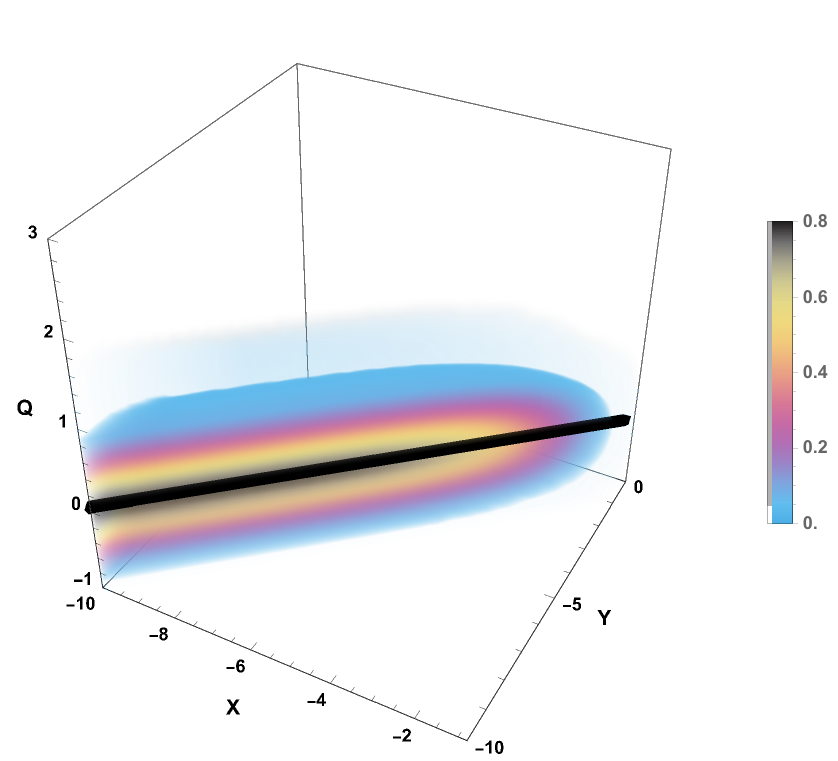}\\   
    \caption{Left: The squared modulus of the wave function $\Psi_e$ Eq.(\ref{eq:gbwfe}) with $\{r_s,\sigma,r_Q\}=\{1,1,0\}$ is shown. The black line is the classical trajectory. Right: The planar cross-section shows that the amplitude maximum traces the steepest-descent contour.} 
    \label{fig:WF3}
    \end{center}
\end{figure}

In the monotonic‑decay scenario, Schwarzschild and Schwarzschild–(anti)‑de Sitter Wheeler–DeWitt solutions satisfy the DeWitt boundary condition at the curvature singularity, realizing singularity avoidance; however, prior analyses achieve this at the cost of an off‑shell growth that makes the wave function unbounded (see Sec. III.D of Ref. \cite{Bouhmadi-Lopez:2019kkt} and Sec. III.A of Ref. \cite{Chien:2023kqw} under the purely incoming horizon condition). In contrast, by reinterpreting \textit{the Schwarzschild black hole as the charge‑neutral limit of the Reissner–Nordstr\"{o}m black hole} and implementing a Gaussian wave with charge fluctuations as in Eq. \eqref{eq:gbwfe}, the neutral‑limit state is normalizable and exhibits strict monotonic decay. This construction furnishes a bounded realization of the monotonic‑decay solution while preserving the DeWitt boundary condition, therefore providing a new perspective on neutralization and refining the late‑time Schwarzschild end‑state picture (cf. Refs. \cite{Bouhmadi-Lopez:2019kkt,Chien:2023kqw}), as illustrated in Fig. \ref{fig:WF3}).

\section{Perspectives on Arrow of Time and Comparison With Other Models\label{sec:AT}}
\begin{figure}[h]
    \begin{center}
    \includegraphics[width=0.30\textwidth]{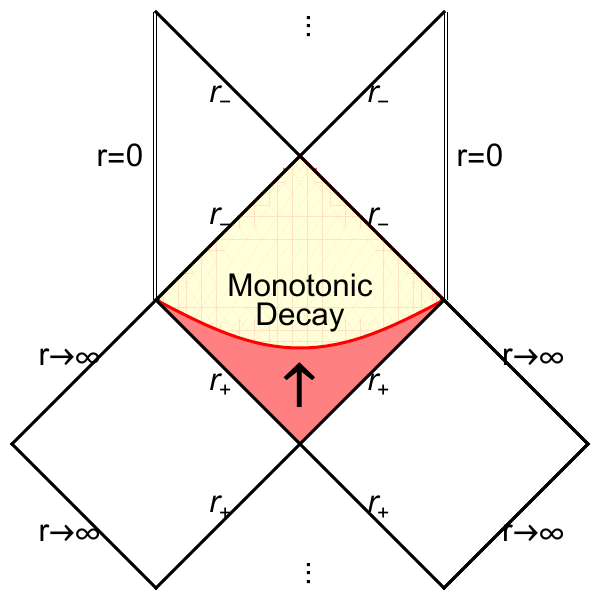}\hspace{0.05 cm}
    \includegraphics[width=0.30\textwidth]{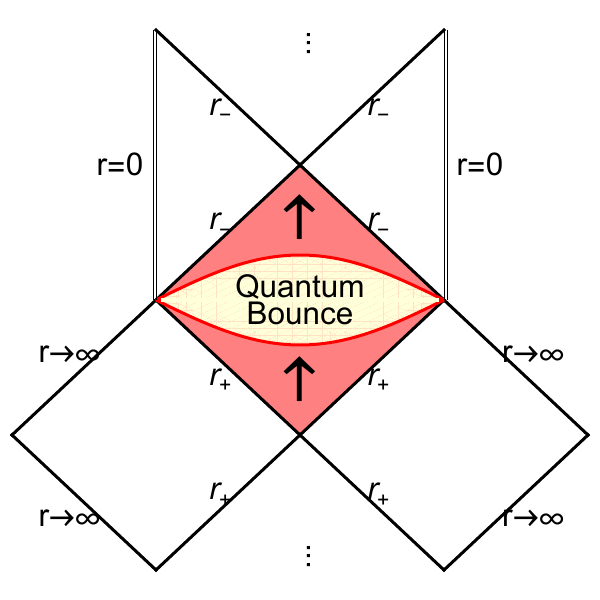}\hspace{0.05 cm}
    \includegraphics[width=0.30\textwidth]{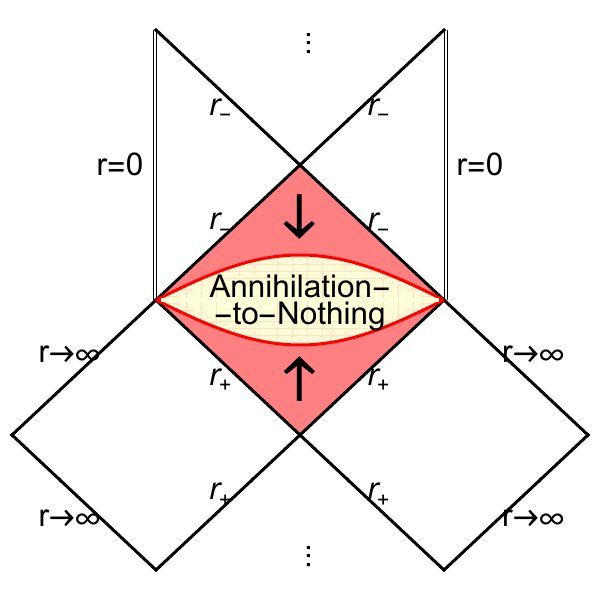}
    \caption{Penrose diagram of Reissner–Nordstr\"{o}m black hole is presented. The arrow is the arrow of time. The waves $\Psi_+$ from $r_+$ and $\Psi_-$ from $r_-$ refer to the red region. The yellow region is the empty space without any waves.} 
    \label{fig:PD}
    \end{center}
\end{figure}

Along the steepest-descent paths of the wave function, one can define an arrow of time. As suggested by the continuity at the classical boundary $r_+$, the wave $\Psi_+$ carries an inward-pointing arrow of time. However, the presence of $\Psi_-$ introduces an ambiguity in the definition of the arrow of time. From a classical perspective, there exists only a single arrow of time pointing inward from $r_+$ to $r_-$. In this case, the wave function may exhibit either monotonic decay or a quantum bounce, as illustrated in the left and center panels of Fig.~\ref{fig:PD}. On the other hand, if two arrows of time are admitted, the interpretation would correspond to two waves annihilating each other, as shown in the right panel of Fig.~\ref{fig:PD}. This is analogous to the ``annihilating-to-nothing'' scenario (see Ref.~\cite{Bouhmadi-Lopez:2019kkt} for further discussion).

In the monotonic‑decay scenario, reinterpreting the Schwarzschild geometry as the charge‑neutral limit of Reissner–Nordstr\"{o}m yields a Wheeler–DeWitt state that is bounded and satisfies the DeWitt boundary condition at the curvature singularity, thus realizing singularity avoidance (see Sec. \ref{sec:NBH}). We further conjecture that the same bounded, monotonic‑decay behaviour extends to Schwarzschild–(anti‑)de Sitter, with the apparent unboundedness reported in Ref. \cite{Chien:2023kqw} attributable to numerical artifacts rather than a physical obstruction. By contrast, for Reissner–Nordstr\"{o}m the monotonic‑decay solution arises naturally and does not exhibit these issues.

For the bounce or annihilation scenarios, Schwarzschild and Schwarzschild–(anti-)de Sitter black holes exhibit a nonzero wave function at the singularity, which may be problematic. In these cases, the DeWitt condition effectively localizes around half the range of the timelike coordinate. Similarly, the wave function in Reissner–Nordstr\"{o}m vanishes over a broader interval of the timelike coordinate $Q$, creating an empty region between $\Psi_+$ and $\Psi_-$ (as shown in the Fig.~\ref{fig:WF2}). We thus conclude that the DeWitt boundary condition localizes near approximately half the extent of the timelike coordinate, yielding either a quantum bounce or an “annihilation-to-nothing” behavior.

Therefore, we conclude that these scenarios—monotonic decay, quantum bounce, and “annihilation-to-nothing”—are generally possible features of the interior of a static black hole; see Table~\ref{tab:BHI}.

\section{Comments on Cauchy Horizon and Mass Inflation\label{sec:CH}}
It is worth emphasizing that analyses of charged black hole interiors are intrinsically tied to the stability properties of the inner (Cauchy) horizon.

Around the inner Cauchy horizon, it has been known that there exists an infinite blue shift \cite{Simpson:1973ua}. This will make the inner horizon unstable. More specifically, if matter fluctuations exist near the inner horizon and an observer measures the matter field fluctuations, the observer will observe exponential growth in local energy or curvature quantities, a phenomenon known as mass inflation \cite{Poisson:1989zz}. Hence, for dynamical situations, there is no guarantee that one can trust the Reissner-Nordstr\"{o}m solution or even that its perturbative analysis will be effective, and complete numerical computations are necessary \cite{Hong:2008mw}.

In this context, our study offers a novel approach to the problem. If we interpret the quantum gravitational wave function as “annihilation-to-nothing” inside the Reissner-Nordstr\"{o}m black hole, one can interpret that both the event horizon and the Cauchy horizon are the causal past of the quantum bouncing point. The infinite blue shift is a problem only if the Cauchy horizon is at the causal future.

Furthermore, if we think that there is only one piece of the incoming wave from the event horizon, then it may indicate that there is no physical structure around the Cauchy horizon. Of course, this is not classically complete, but by choosing a suitable boundary condition of the quantum gravitational wave function, we can smoothly terminate physical geometries inside the horizon. This choice of boundary condition suppresses the mass inflation problem by removing the interior geometry of charged black holes in a quantum-gravitational manner.

Can this interpretation be generalized for all black holes that have inner Cauchy horizons, e.g., Kerr black holes or some regular black holes (see, for example, \cite{Brown:2011tv})? At this moment, we cannot be sure about this, but the quantum gravitational approach might shed some light on resolving the inner horizon instability issue. We leave this interesting topic for future work.

\section{Conclusion\label{sec:con}}
In this work, we have developed a quantization of the interior geometry of static and spherically symmetric black holes within the Einstein-Maxwell-$\Lambda$ framework, focusing in particular on the Reissner-Nordstr\"{o}m case. Starting from the full Einstein-Maxwell-$\Lambda$ action, we derived the reduced Lagrangian, the canonical Hamiltonian, and the corresponding Wheeler-DeWitt equation in the three-dimensional superspace spanned by $\{X,Y,Q\}$. The classical analysis was characterized via the constraint surface relation and explicit parametric solutions between the event and Cauchy horizons, revealing distinct asymptotic behavior at each horizon and a charge-dependent spreading of trajectories.

The quantum analysis, implemented via separation of variables, produced bounded solutions in terms of modified Bessel functions, whose asymptotic behavior enforces exponential damping away from the classical locus. We examined the steepest–descent structure of the wave function, which is sharply localized along on-shell trajectories. Within the Gaussian approximation, the wave packet is strongly localized in $Q$-space set by a standard deviation $\sigma$. In practice, this broadening choice of $\sigma$ leads to an undesirable overlap between the $\Psi_{+}$ and $\Psi_{-}$ components. This overlap should not be interpreted as a physical feature of the Reissner–Nordstr\"{o}m interior; rather, it arises as a numerical issue originating from the stability-localization trade-off in the choice of $\sigma$. Furthermore, by viewing the Schwarzschild geometry as the charge‑neutral limit of the Reissner–Nordstr\"{o}m family, the associated Wheeler–DeWitt state becomes bounded and satisfies the DeWitt boundary condition at the curvature singularity, achieving singularity avoidance.

Special attention was given to the role of boundary conditions at the event and Cauchy horizons, resulting in two physically distinct scenarios: (i) A single inward‑propagating mode from $r_{+}$ fixes a single arrow of time and yields monotonic decay. (ii) A superposition of contributions from both horizons yields: with a single time arrow, a quantum bounce; with counter‑propagating components (two time arrows), interference leading to “annihilation‑to‑nothing”. By analogy with the Schwarzschild interior, the Reissner–Nordstr\"{o}m case exhibits a similar decay, bounce, or annihilation localized at a timelike $Q$ coordinate, suggesting that such features are generic to the interiors of static black holes, as shown in Table~\ref {tab:BHI}.

Analyses of charged black‑hole interiors are inseparable from the instability of the inner Cauchy horizon, where infinite blueshift drives mass inflation and undermines perturbative control. This work advances a complementary quantum‑gravitational suppression: (i) Interpreting the interior Wheeler–DeWitt state as an “annihilation‑to‑nothing” bounce that places both the event and Cauchy horizons in the causal past of the quantum turning point. (ii) Imposing a purely incoming boundary condition at the event horizon then smoothly terminates the interior geometry before a physical Cauchy horizon forms, thus eliminating the mass-inflation region and preventing the associated instabilities. Whether these mechanisms extend to other inner‑horizon spacetimes, including Kerr and regular black holes, remains an open question and a promising avenue for future investigation.

While the general features summarized in Table~\ref{tab:BHI} for small $\Lambda$ and $r_Q$ are expected, it is crucial to underscore that these outcomes are contingent upon the boundary conditions at the inner horizon and the singularity. Acknowledging that the Kantowski-Sachs metric may not fully capture the global properties of the quantized black hole, our observations remain tentative: Schwarzschild and Schwarzschild–(anti-)de Sitter geometries appear to favor monotonic decay, whereas Reissner–Nordstr\"{o}m black holes manifest both monotonic decay and 'annihilation-to-nothing' channels.

A persistent challenge lies in consistently matching these interior results to the regions outside the event horizon and inside the Cauchy horizon, going beyond the specific boundary prescriptions adopted in this study. A robust resolution might involve employing Kruskal-Szekeres coordinates, as demonstrated in Ref.~\cite{Lin:2025jmz}; however, this approach introduces nontrivial metric parameters, a complexity we reserve for future investigation.

Overall, this framework unifies the classical and quantum descriptions of black-hole interiors, highlights the influence of electromagnetic gauge field on trajectories and localization, and indicates that monotonic decay, quantum bounce, or annihilation is possibly a generic feature of static black holes. Furthermore, the wave function of the Schwarzschild black hole, obtained as the charge-neutral limit of the Reissner–Nordstr\"{o}m black hole, exhibits a monotonically decaying behavior and is no longer unbounded. The fact that we can provide three interpretations, where wave packets can be bounded for all cases, is a new observation of this Reissner–Nordstr\"{o}m black hole. The choice of boundary condition can suppress the instability issue of the inner horizon, as one can physically remove the inside structure or locate the inner horizon as a causal past. Still, we cannot be sure whether we can generalize this rescue to highly dynamic situations, but we believe that it can open a window to resolving the inner horizon instability issue. For further investigations, we leave the topic for future projects.

\appendix
\section{Numerical Instability of Gaussian Wave}\label{appendix:ni}
\begin{figure}[!h]
    \centering
    \includegraphics[width=0.45\textwidth]{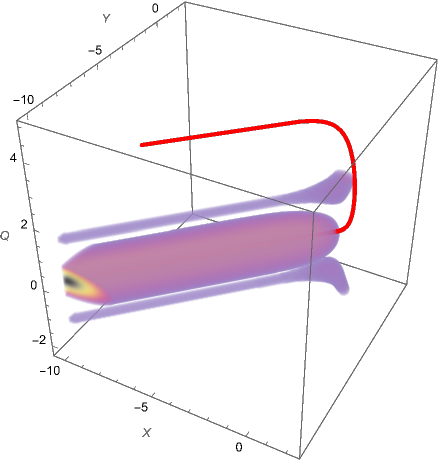}\hspace{0.05 cm}
    \includegraphics[width=0.45\textwidth]{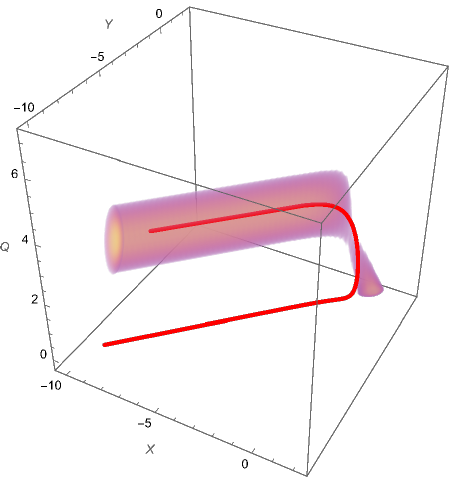}\\
    \caption{Numerical integration profiles for the parameter set $\{r_Q, \sigma, \delta\} = \{0.2, 1, 0.6\}$, evaluated within the boundary range $[-(\lambda_0 + 3\delta), \lambda_0 + 3\delta]$. The left panel displays the wavepacket $\Psi_+$, while the right panel corresponds to $\Psi_-$.} 
    \label{fig:App}
\end{figure}

To mitigate the numerical instabilities associated with the Gaussian wavepacket in Eq.~(\ref{eq:gbwf}), we modify the profile function $g(\lambda)$ in Eq.~(\ref{eq:gw}) as follows:
\begin{align}
g(\lambda) &= \exp\left[ -\left( \frac{|\lambda| - \lambda_0}{\delta} \right)^2 \right] \exp\left[ -2 \left( Y - X + \ln Q - \ln\left(\frac{r_q}{r_s}\right) \right)^2 \right].
\end{align}
The secondary exponential factor enforces localization along the classical trajectory defined in Eq.~(\ref{eq:al}) for $\Psi_-$. As illustrated in Fig.~\ref{fig:App}, this specific choice constrains the width of the wavepacket $\Psi_-$, effectively suppressing numerical overlap. While this approach necessitates an cutoff and careful fine-tuning of the amplitude, it ensures a stable and computationally tractable evolution of the wavepacket.

\begin{acknowledgments}
CHC is supported by the Grant Agency of the Czech Republic under project no. GA24-10887S. DY is supported by the National Research Foundation of Korea (Grant No. 2021R1C1C1008622, No. 2021R1A4A5031460). GT is supported by the National Research Foundation of Korea (NRF), funded by the Ministry of Education (Grant No. NRF-2022R1I1A1A01053784 and NRF-2021R1A2C1005748).

\end{acknowledgments}


\begin{thebibliography}{99}
\bibitem{Hawking:1970zqf}
S.~W.~Hawking and R.~Penrose,
Proc. R. Soc. A \textbf{314}, 529 (1970).

\bibitem{Hawking:1976ra}
S.~W.~Hawking,
Phys. Rev. D \textbf{14}, 2460 (1976).

\bibitem{Boehmer:2007ket}
C.~G.~Boehmer and K.~Vandersloot,
Phys. Rev. D \textbf{76} (2007), 104030
[arXiv:0709.2129 [gr-qc]].

\bibitem{Modesto:2006mx}
L.~Modesto,
Adv. High Energy Phys. \textbf{2008} (2008), 459290
[arXiv:gr-qc/0611043 [gr-qc]].

\bibitem{Zhang:2023yps}
X.~Zhang,
Universe \textbf{9} (2023) no.7, 313
[arXiv:2308.10184 [gr-qc]].

\bibitem{Chew:2023upu}
X.~Y.~Chew and D.~Yeom,
J. Korean Phys. Soc. \textbf{85} (2024) no.12, 1050-1061
[arXiv:2308.09225 [gr-qc]].

\bibitem{Carballo-Rubio:2024dca}
R.~Carballo-Rubio, F.~Di Filippo, S.~Liberati and M.~Visser,
Phys. Rev. Lett. \textbf{133} (2024) no.18, 181402
[arXiv:2402.14913 [gr-qc]].

\bibitem{Cardoso:2017soq}
V.~Cardoso, J.~L.~Costa, K.~Destounis, P.~Hintz and A.~Jansen,
Phys. Rev. Lett. \textbf{120} (2018) no.3, 031103
[arXiv:1711.10502 [gr-qc]].

\bibitem{Bouhmadi-Lopez:2019kkt}
M.~Bouhmadi-L{\'o}pez, S.~Brahma, C.~Y.~Chen, P.~Chen and D.~Yeom,
JCAP \textbf{11} (2020), 002
[arXiv:1911.02129 [gr-qc]].

\bibitem{Brahma:2021xjy}
S.~Brahma, C.~Y.~Chen and D.~Yeom,
Eur. Phys. J. C \textbf{82} (2022) no.9, 772
[arXiv:2108.05330 [gr-qc]].

\bibitem{Yeom:2021bpd}
D.~Yeom,
[arXiv:2105.00066 [gr-qc]].

\bibitem{Chien:2023kqw}
C.~H.~Chien, G.~Tumurtushaa and D.~Yeom,
Phys. Rev. D \textbf{108} (2023) no.2, 023530
[arXiv:2302.00254 [gr-qc]].


\bibitem{Singh:2024kcn}
H.~Singh and M.~K.~Nandy,
Eur. Phys. J. C \textbf{84} (2024) no.7, 700.

\bibitem{Kan:2022ism}
N.~Kan, T.~Aoyama and K.~Shiraishi,
Class. Quant. Grav. \textbf{40} (2023) no.16, 165006
[arXiv:2209.14527 [gr-qc]].

\bibitem{Kantowski:1966te}
R.~Kantowski and R.~K.~Sachs,
J. Math. Phys. \textbf{7} (1966), 443.

\bibitem{Poisson:1989zz}
E.~Poisson and W.~Israel,
Phys. Rev. Lett. \textbf{63}, 1663-1666 (1989).

\bibitem{McMaken:2023tft}
T.~McMaken and A.~J.~S.~Hamilton,
Phys. Rev. D \textbf{107}, no.8, 085010 (2023)
[arXiv:2301.12319 [gr-qc]].

\bibitem{Burko:1997zy}
L.~M.~Burko,
Phys. Rev. Lett. \textbf{79}, 4958-4961 (1997)
[arXiv:gr-qc/9710112 [gr-qc]].

\bibitem{Cavaglia:1994yc}
M.~Cavaglia, V.~de Alfaro and A.~T.~Filippov,
Int. J. Mod. Phys. D \textbf{4}, 661 (1995).

\bibitem{Garcia-Compean:2001jxk}
H.~Garcia-Compean, O.~Obregon and C.~Ramirez,
Phys. Rev. Lett. \textbf{88}, 161301 (2002).

\bibitem{Lopez-Dominguez:2006rou}
J.~C.~Lopez-Dominguez, O.~Obregon, M.~Sabido and C.~Ramirez,
Phys. Rev. D \textbf{74}, 084024 (2006).

\bibitem{Bastos:2007bg}
C.~Bastos, O.~Bertolami, N.~Costa Dias and J.~Nuno Prata,
Phys. Rev. D \textbf{78}, 023516 (2008).

\bibitem{Moniz:1997ur}
P.~V.~Moniz,
Mod. Phys. Lett. A \textbf{12} (1997), 1491-1505
[arXiv:gr-qc/9709080 [gr-qc]].

\bibitem{Garattini:2008qs}
R.~Garattini,
Phys. Lett. B \textbf{666} (2008), 189-192
[arXiv:0807.0082 [gr-qc]].

\bibitem{Blacker:2023ezy}
M.~J.~Blacker and S.~Ning,
JHEP \textbf{12} (2023), 002
[arXiv:2308.00040 [hep-th]].

\bibitem{gradshteyn2014table}
I. S. Gradshteyn and I. M. Ryzhik
\textit{Tables of Integrals, Series, and Products}, (Academic Press, New York, 2014).

\bibitem{Simpson:1973ua}
M.~Simpson and R.~Penrose,
Int. J. Theor. Phys. \textbf{7}, 183-197 (1973).

\bibitem{Gibbons:1975kk}
G.~W.~Gibbons,
Commun. Math. Phys. \textbf{44}, 245-264 (1975)
doi:10.1007/BF01609829

\bibitem{Page:1977um}
D.~N.~Page,
Phys. Rev. D \textbf{16}, 2402-2411 (1977)
doi:10.1103/PhysRevD.16.2402

\bibitem{Iso:2006wa}
S.~Iso, H.~Umetsu and F.~Wilczek,
Phys. Rev. Lett. \textbf{96}, 151302 (2006)
doi:10.1103/PhysRevLett.96.151302
[arXiv:hep-th/0602146 [hep-th]].

\bibitem{Gong:2019aqa}
Y.~Gong, Z.~Cao, H.~Gao and B.~Zhang,
Mon. Not. Roy. Astron. Soc. \textbf{488}, no.2, 2722-2731 (2019)
doi:10.1093/mnras/stz1904
[arXiv:1907.05239 [gr-qc]].

\bibitem{Hong:2008mw}
S.~E.~Hong, D.~Hwang, E.~D.~Stewart and D.~Yeom,
Class. Quant. Grav. \textbf{27}, 045014 (2010)
[arXiv:0808.1709 [gr-qc]].

\bibitem{Brown:2011tv}
E.~G.~Brown, R.~B.~Mann and L.~Modesto,
Phys. Rev. D \textbf{84}, 104041 (2011)
[arXiv:1104.3126 [gr-qc]].

\bibitem{Lin:2025jmz}
W.~C.~Lin, D.~h.~Yeom and D.~Stojkovic,
[arXiv:2510.20799 [gr-qc]].

\end{thebibliography}
\end{document}